\documentclass[envcountsame,envcountsect,undated,nolinenumbers]{lnthi}

\usepackage{graphics,amsmath}

\title{Rank-Select Indices Without Tears}

\author{Tim Baumann\inst{1} and Torben Hagerup\inst{2}}
\institute{Institut f\"ur Mathematik, Universit\"at Augsburg, 86135
Augsburg, Germany\\\email{tim@timbaumann.info} \and
Institut f\"ur Informatik, Universit\"at Augsburg, 86135
Augsburg, Germany
\email{hagerup@informatik.uni-augsburg.de}}

\setpagewiselinenumbers

\def\Tvn#1{\hbox{\textit{#1\/}}}
\def\Tfloor#1{\lfloor #1\rfloor}
\def\Tceil#1{\lceil #1\rceil}
\def\TbbbN{\mathbb{N}}
\def\TbbbZ{\mathbb{Z}}
\gdef\Tsub#1{_{\mbox{\scriptsize #1}}}
\def\ms#1{(\mskip-3mu(#1)\mskip-3mu)}
\def\simple#1{\Tvn{supp}(#1)}
\def\Tangle#1{\langle#1\rangle}
\def\rank{\Tvn{rank}}
\def\select{\Tvn{select}}

\begin{document}
  \overfullrule=5pt

\maketitle{}

\begin{abstract}
A \emph{rank-select index} for a
sequence $B=(b_1,\ldots,b_n)$ of $n$ bits,
where $n\in\TbbbN=\{1,2,\ldots\}$,
is a data structure that, if
provided with an operation
to access $\Theta(\log n)$ arbitrary
consecutive bits of $B$ in constant time
(thus $B$ is stored outside of the data structure),
can compute
$\rank_B(j)=\sum_{i=1}^j b_i$ for given
$j\in\{0,\ldots,n\}$ and
$\select_B(k)=\min\{j\in\TbbbN\mid\rank_B(j)\ge k\}$
for given $k\in\{1,\ldots,\sum_{i=1}^n b_i\}$.
We describe a new rank-select index that, like
previous rank-select indices, occupies
$O({{n\log\log n}/{\log n}})$ bits and
executes $\rank$ and $\select$ queries in constant time.
Its derivation is intended to be particularly
easy to follow and largely free of
tedious low-level detail,
its operations are given by
straight-line code, and we show that it can be
constructed in $O({n/{\log n}})$ time.
\end{abstract}

\section{Introduction}

When $S$ is a finite multiset of integers and
$j\in\TbbbZ=\{\ldots,-1,0,1,\ldots\}$,
we write $\rank_S(j)$ for the rank of $j$ in $S$, i.e.,
$\rank_S(j)=|\{i\in S:i\le j\}|$.
Moreover, for each $k\in\{1,\ldots,|S|\}$,
$\select_S(k)=\min\{j\in\TbbbZ:\rank_S(j)\ge k\}$.
If the elements of $S$ are arranged in
nondecreasing order in positions
$1,\ldots,|S|$, then $\rank_S(j)$ is the
largest position of an element $\le j$
(0 if there is no such element), for $j\in\TbbbZ$,
and $\select_S(k)$ is the element in position~$k$,
for $k\in\{1,\ldots,|S|\}$.

The operations $\rank$ and $\select$ are also defined
for bit sequences.
If $B=(b_1,\ldots,b_n)$ is a sequence of $n$ bits,
for some $n\in\TbbbN=\{1,2,\ldots\}$, then
$\rank_B(j)=\sum_{i=1}^j b_i$ for $j\in\{0,\ldots,n\}$ and,
again,
$\select_B(k)=\min\{j\in\TbbbN:\rank_B(j)\ge k\}$
for $k\in\{1,\ldots,\sum_{i=1}^n b_i\}$.
The connection between the two definitions is close:
If a simple set $S$ is a subset of
$\{a,\ldots,a+n-1\}$ for some known $a\in\TbbbZ$
and $n\in\TbbbN$, $S$ can be represented via the
bit sequence $(b_1,\ldots,b_n)$ with
$b_i=1\Leftrightarrow a-1+i\in S$,
for $i=1,\ldots,n$.
In this case we say that $S$ is given by its
\emph{bit-vector representation} over the
universe $\{a,\ldots,a+n-1\}$ or
with \emph{offset} $a$ and
\emph{span}~$n$.
Clearly $\rank_S(j)=\rank_B(j-(a-1))$ for
$j\in\{a-1,\ldots,a-1+n\}$
and $\select_S(k)=a-1+\select_B(k)$
for $k\in\{1,\ldots,|S|\}$.
Answering $\rank$ and $\select$ queries about a
simple set of integers therefore reduces to
answering $\rank$ and $\select$ queries about
its bit-vector representation with some known offset and span.

A (static) \emph{rank-select structure}
for a sequence $B=(b_1,\ldots,b_n)$ of $n$ bits, for some $n\in\TbbbN$,
is a data structure capable of
returning $\rank_B(j)$ for arbitrary given
$j\in\{0,\ldots,n\}$ and $\select_B(k)$ for arbitrary
given $k\in\{1,\ldots,\sum_{i=1}^n b_i\}$.
A data structure that can answer the same queries, but
only if provided with an operation
to access $\Theta(\log n)$ arbitrary
consecutive bits of $B$ in constant time
(thus $B$ is stored outside of the data structure),
is known as a \emph{rank-select index} for~$B$.
We call $B$ the \emph{client sequence} of a
rank-select structure or index for~$B$.
Rank-select structures and indices are of fundamental
importance in space-efficient computing,
have been studied extensively since the
1970s, and have many
and diverse applications.
Elias~\cite{Eli74} considered the representation of
multisets of integers in the context of
data retrieval.
For a multiset $S$, his \emph{direct}
or \emph{table-lookup} question corresponds to
$\select_S$, and his \emph{inverse}
question is closely related to $\rank_S$.
Jacobson, who introduced the terms
$\rank$ and $\select$~\cite{Jac88,Jac89},
used rank-select structures to represent trees
and graphs in little space while still permitting
their efficient traversal.
Along the way, he solved the problem
of finding matching parentheses in a balanced
sequence of parentheses, again with rank-select
structures as crucial components of the
overall data structure.
Rank-select structures and indices
have also found applications
in areas such as string processing~\cite{GroGV03},
computational geometry~\cite{MakN07}
and graph algorithms~\cite{HagKL17}.

Jacobson designed a rank-select index
for bit sequences of length~$n\in\TbbbN$
that occupies
$O({{n\log\log n}/{\log n}})$ bits and
answers $\rank$ queries in constant time.
While he was unable to obtain a constant-time
$\select$ operation, this was remedied by
Clark~\cite{Cla96}, at the price of a somewhat
higher space bound.
From now on we will be interested only in
rank-select structures and indices that answer all
queries in constant time; for ease of discussion,
consider this property to be part of their definition.

A rank-select index that
uses $O({{n\log\log n}/{\log n}})$ bits
was described by
Raman, Raman and Satti~\cite[Lemma 4.1]{RamRS07}.
A matching lower bound of $\Omega({{n\log\log n}/{\log n}})$
on the number of bits needed by a rank-select
index that accesses only $O(\log n)$ bits of
its client sequence during the processing of a query
was proved by Golynski~\cite{Gol07}.
Thus a rank-select structure for a sequence $B$
of $n$ bits
that consists of $B$ plus a suitable index
must occupy $n+\Omega({{n\log\log n}/{\log n}})$ bits.
While this is a natural way of organizing a
rank-select structure, P\v atra\c scu~\cite{Pat08}
proved the interesting fact that
there are smaller rank-select structures that do not
store $B$ in its ``raw'' form.

In this paper we describe another
$O({{n\log\log n}/{\log n}})$-bit rank-select index.
While previous descriptions of
rank-select structures and indices
abund with ad-hoc and rather tedious low-level detail,
we aim for a more systematic and high-level approach
based largely on pictures that
leads to the optimal result with
little effort on the part of the reader.
Our rank-select index offers the first
$\select$ operation that can be formulated as a piece
of straight-line code, i.e., its implementation
is free of tests and branching
(in one place, fulfilling this promise involves
a small amount of ``cheating'',
as will be explained later).
We also consider the problem of efficient
construction of rank-select indices,
an aspect that was ignored in much previous research
but is essential to many applications.
Our main result is the following:

\begin{theorem}
\label{thm:main}
For every $n\in\TbbbN$ and for every
sequence $B$ of $n$ bits,
given in the form of a stream
of $O({n/{\log n}})$ chunks of
$O({{n\log\log n}/{\log n}})$ consecutive bits each,
a rank-select index for $B$ that executes
$\rank_B$ and $\select_B$ in constant time and
occupies $O({{n\log\log n}/{\log n}})$ bits can
be constructed in $O({n/{\log n}})$ time
using $O({{n\log\log n}/{\log n}})$ bits
of working memory.
\end{theorem}

The theorem insists that the client sequence~$B$
be provided in several chunks because
the available working space does not allow
us to store $B$ in its entirety.
Typically $B$ would be provided in
$\Theta({n/{\log n}})$ chunks of
$O(\log n)$ bits each.
If $B$ is stored in random-access read-only
memory, of course, it is trivial to produce the
necessary chunks, but the theorem implies that
the construction of the rank-select index does
not require random access to $B$ and can make
do with a single pass over~$B$.

Our model of computation is a word
RAM~\cite{AngV79,Hag98} with a word length
of $w=\Omega(\log n)$ bits, where $w$ is assumed
large enough to allow all memory words
in use to be addressed.
The word RAM has constant-time operations for
addition, subtraction and multiplication
modulo $2^w$, division with truncation
($(x,y)\mapsto\Tfloor{{x/y}}$ for $y>0$),
left shift modulo $2^w$
($(x,y)\mapsto (x\ll y)\bmod 2^w$,
where $x\ll y=x\cdot 2^y$),
right shift
($(x,y)\mapsto x\gg y=\Tfloor{{x/{2^y}}}$),
and bitwise Boolean operations
($\textsc{and}$, $\textsc{or}$ and $\textsc{xor}$
(exclusive or)).

\section{Ingredients of the New Rank-Select Index}

Our overall approach, shared with earlier solutions,
is to break down a given instance of the
rank-select problem, i.e., the problem
of answering $\rank$ and $\select$ queries
for a given bit sequence or multiset,
into still smaller instances,
eventually arriving
at instances so tiny that they can
be solved by brute force, i.e., table lookup.
We provide a bottom-up description, proceeding from
table lookup via basic reductions of instances
of the rank-select problem to simpler instances
and ending with the complete rank-select index
that reduces $\rank$ and $\select$ queries about
the client sequence all the way to table lookup.
We prefer to phrase much of the discussion in terms
of (multi)sets rather than bit sequences.
The tables needed by the rank-select index
and their computation are
discussed in the next subsection.

\subsection{Table Lookup}
\label{subsec:tables}

This subsection describes three different
variants, denoted T1--T3, of the table-lookup method,
as applied to the rank-select problem.
In our applications, the parameters $N$ and $M$
for variants T2 and T3
will be so small as to render negligible
the space occupied by the tables and the
time needed to compute them.

\paragraph*{T1.}
In order to answer $\rank_S$ and $\select_S$
queries about one particular subset $S$ of
$\{1,\ldots,N\}$, where $N\in\TbbbN$ is known,
we can simply store
a table of
$\rank_S(j)$ for $j=0,\ldots,N$
and $\select_S(k)$ for $k=1,\ldots,|S|$
and answer a query by returning
an appropriate table entry.
The number of bits needed is $O(N\log N)$,
and the table can be computed in $O(N)$ time
from a bit-vector representation of~$S$.

\paragraph*{T2.}
If the goal is to answer
$\rank_S$ and $\select_S$ queries,
where $S$ now is also
specified in the query
and can be an arbitrary subset
of $\{1,\ldots,N\}$, we can create
a subtable as for variant T1
for each of the $2^N$ possible subsets~$S$
and store the $2^N$ subtables, each indexed
by the bit-vector representation
over $\{1,\ldots,N\}$
of the corresponding set~$S$,
in a table of $O(2^N N\log N)$ bits whose
computation takes $O(2^N N)$ time.

\paragraph*{T3.}
If $S$ is a variable subset of $\{1,\ldots,N\}$
but known to be of
size at most~$M$ for some given $M\in\TbbbN$,
$S$ can be represented
as an $M$-tuple of integers in $\{1,\ldots,N\}$
by first listing the elements of $S$ and
then, if $|S|<M$, repeating the last element.
Each of the $M$ integers can in turn be represented
in $\Tceil{\log_2\! N}$ bits.
Since $2^{\Tceil{\log_2\! N}}\le 2 N$,
this gives us an alternative to variant~T2
with a table of $O((2 N)^{M+1}\log N)$ bits that
can be computed in $O((2 N)^{M+1})$ time.

\subsection{Three Basic Reductions}
\label{subsec:basic}

\begin{figure}[!ht]
\begin{center}
\includegraphics{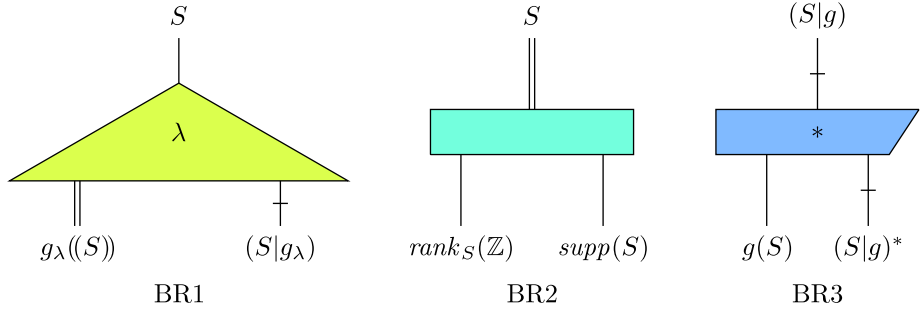}
\end{center}
\caption{The three basic reductions BR1--BR3.}
\label{fig:basic}
\end{figure}

Let $g$ be a nondecreasing
function from $\TbbbZ$ to $\TbbbZ$
(informally, the \emph{grouping function})
with the property that
$g^{-1}(q)=\{i\in\TbbbZ\mid g(i)=q\}$
is finite for all $q\in\TbbbZ$ and let
$S$ be a finite multiset of integers.
While $g(S)$ as usual denotes the simple
set $\{q\in\TbbbZ\mid S\cap g^{-1}(q)\not=\emptyset\}$,
we write $g\ms{S}$ for the \emph{multiset}
$\{g(i)\mid i\in S\}$, in which each $q\in\TbbbZ$
occurs with multiplicity
$|S\cap g^{-1}(q)|$.
For $j\in\TbbbZ$, it is clear that,
with $q=g(j)$,
\begin{equation}
\rank_S(j)=\rank_{g\ms{S}}(q-1)+\rank_{S\cap g^{-1}(q)}(j),
\label{eq:1}
\end{equation}
since the terms $\rank_{g\ms{S}}(q-1)$ and
$\rank_{S\cap g^{-1}(q)}(j)$ count the elements $i$ of
$\{i\in S:i\le j\}$ with $g(i)<q$
and with $g(i)=q$, respectively.
Furthermore, for $k\in\{1,\ldots,|S|\}$,
$g(\select_S(k))=\select_{g\ms{S}}(k)$ and therefore,
with $q=\select_{g\ms{S}}(k)$,
\begin{equation}
\select_S(k)=\select_{S\cap g^{-1}(q)}(k-\rank_{g\ms{S}}(q-1)),
\label{eq:2}
\end{equation}
since $\rank_{g\ms{S}}(q-1)$ again
is the number of elements
$i\in S$ with $g(i)<q$, so that, if $S$
is presented in sorted order in $|S|$ positions,
$\select_S(k)$
is the element in position $k-\rank_{g\ms{S}}(q-1)$
among the elements with the same value
under $g$ as itself, i.e., within $S\cap g^{-1}(q)$.
Let us use $(S|g)$ as a convenient notation for the function
$(g_{|S})^{-1}$ that maps each
$q\in\TbbbZ$ to $S\cap g^{-1}(q)$.
Then answering $\rank$ and $\select$ queries about $S$
in constant time
reduces
to answering $\rank$ and $\select$ queries
about $g\ms{S}$ and about
values of $(S|g)$ in constant time.
For brevity, we express this by saying that
$S$ reduces to $g\ms{S}$ and $(S|g)$.
We will use this only with $g=g_\lambda$
for some $\lambda\in\TbbbN$, where
$g_\lambda(i)=\Tfloor{i/\lambda}$ for
all $i\in\TbbbZ$,
and call $\lambda$ the \emph{parameter} of the reduction.
Variations of this reduction have been used since
the early days of rank-select indices~\cite{Jac88,Jac89}.
We call it BR1 (``basic reduction~1'')
and denote it symbolically with
a triangular shape, as shown in the left
subfigure of Fig.~\ref{fig:basic}:
$S$, at the apex of the triangle, reduces to
$g_\lambda\ms{S}$ and
$(S|g_\lambda)$
at its base.
The double line serves as a reminder that $g_\lambda\ms{S}$
in general is a multiset even if $S$ is not, and a
bar through the line to $(S|g_\lambda)$ indicates that
$(S|g_\lambda)$ is a set-valued function rather than just a set.
We refrain from connecting $S$ to the triangle with a
double line because we will use the reduction
only for simple sets~$S$.

Denote by $\simple{S}$ the support of the
multiset $S$, i.e., the simple set
that contains exactly
the same values as $S$, but each value only once, and
let $\rank_S(\TbbbZ)$ be the image of the $\rank_S$ function.
For example, with $S = \{ 1, 1, 2, 4, 4, 4 \}$ we have
$\simple{S} = \{ 1, 2, 4 \}$ and
$\rank_S(\TbbbZ) = \{ 0, 2, 3, 6 \}$.
A second reduction is given by the formulas
\begin{align}
\rank_S(j) \label{eq:3}
&= \select_{\rank_S(\TbbbZ)}(\rank_{\simple{S}}(j)+1)\mbox{\quad and} \\[0.15em]
\select_S(k) \label{eq:4}
&= \select_{\simple{S}}(\rank_{\rank_S(\TbbbZ)}(k-1)),
\end{align}
for $j\in\TbbbZ$ and $k\in\{1,\ldots,|S|\}$.
To see the validity of the first formula, whose
origins can be traced back to
Fano~\cite[Step~2]{Fan71}, note that
the $(q+1)$st smallest element of $\rank_S(\TbbbZ)$, for
$q=0,\ldots,|\simple{S}|$, is the total
number of occurrences
in $S$ of the $q$ smallest distinct values in~$S$.
With $q=\rank_{\simple{S}}(j)$, this is precisely
$\rank_S(j)$.
The second formula is implied by the following
observation:
If $S$, presented in sorted order in $|S|$ positions,
is thought of as partitioned into maximal \emph{ranges}
of occurrences of the same value, then
$\rank_{\rank_S(\TbbbZ)}(k-1)$ is one more than the
number of ranges that end strictly before the
$k$th position.
Thus $S$
also reduces to
$\rank_S(\TbbbZ)$ and $\simple{S}$.
We call this reduction BR2 and
depict it as shown in
the middle subfigure of Fig.~\ref{fig:basic}.

Denote by $(S|g)^*$ the function
defined on $\{1,\ldots,|g(S)|\}$
that maps $q$ to
$(S|g)(\select_{g(S)}(q))$,
for $q=1,\ldots,|g(S)|$.
Informally, if $(S|g)$ is thought of as a list
of subsets of $S$, then $(S|g)^*$ is the sublist
that contains only the nonempty subsets.
Clearly, for $q\in\TbbbZ$,
\begin{equation}
(S|g)(q)=
\begin{cases}
(S|g)^*(\rank_{g(S)}(q)),&\mbox{ if }q\in g(S),\cr
 \emptyset,&\mbox{ otherwise}.\cr
\end{cases}
\label{eq:5}
\end{equation}
Since we can test whether $q\in g(S)$
by evaluating
$\rank_{g(S)}(q)-\rank_{g(S)}(q-1)$,
which is 1 if $q\in g(S)$ and 0 otherwise,
$(S|g)$ (i.e., $(S|g)(q)$ for each
$q\in\TbbbZ$) reduces to $g(S)$ and $(S|g)^*$.
This third and last basic
reduction, BR3, is depicted in
the right subfigure of Fig.~\ref{fig:basic}.

\subsection{Two Combined Reductions}
\label{subsec:combined}

\begin{figure}[!ht]
\begin{center}
\includegraphics{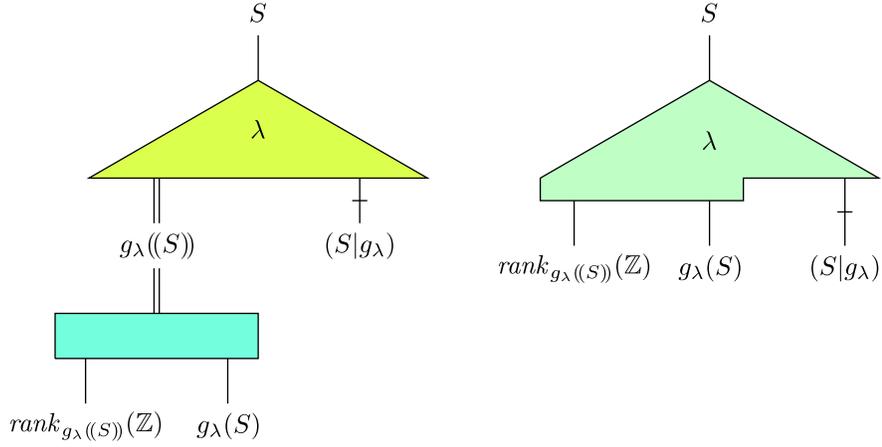}
\end{center}
\caption{The first combined reduction CR1.
Left: Internal structure.
Right: Pictorial representation.}
\label{fig:combined1}
\end{figure}

\begin{figure}[!ht]
\begin{center}
\includegraphics{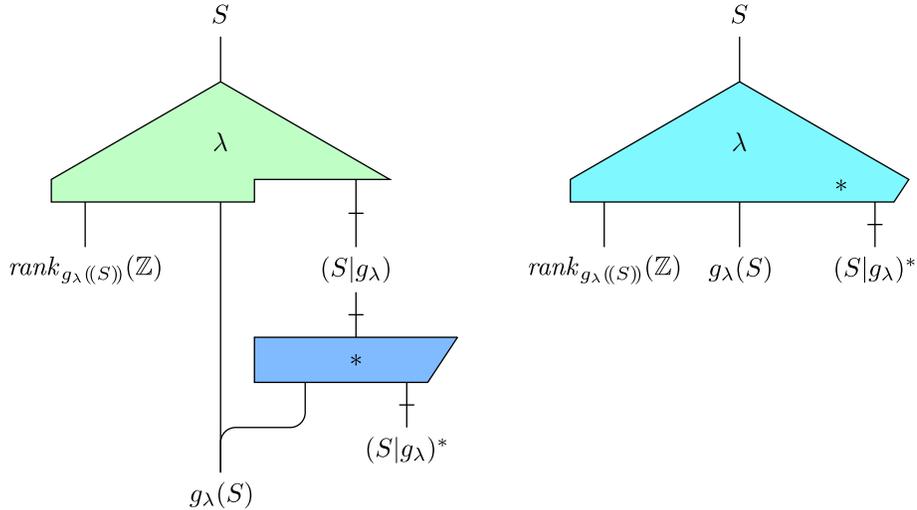}
\end{center}
\caption{The second combined reduction CR2.
Left: Internal structure.
Right: pictorial representation.}
\label{fig:combined2}
\end{figure}

We can combine
BR1 and BR2
as illustrated in Fig.~\ref{fig:combined1}.
This yields a reduction, CR1, of
$S$ to
$\rank_{g_\lambda\ms{S}}(\TbbbZ)$,
$\simple{g_\lambda\ms{S}}=g_\lambda(S)$,
and $(S|g_\lambda)$.
Incorporating also BR3,
we obtain a
second combined reduction, CR2, shown in
Fig.~\ref{fig:combined2}.

In the concrete
rank-select index, sets of integers are represented
as bit vectors with convenient offsets and spans.
The offsets and spans are not stored
with every set, but calculated
in parallel with the application of reductions
according to the following rules:

The client sequence $B$, prefixed by a \texttt{0}
(see later), is viewed as representing
a set with offset 0 and span $|B|+1$.
Recursively,
if a set $S$ has
offset $a$ and span $n$ and $m=|S|$,
then
\begin{itemize}
\item
$\rank_{g_\lambda\ms{S}}(\TbbbZ)$ has offset~0,
span $m+1$ and size at most $\Tceil{n/\lambda}+1$.
\item
$g_\lambda(S)$ has offset $g_\lambda(a)$,
span $\Tceil{n/\lambda}+1$ and size at most~$m$.
\item
$(S|g_\lambda)(q)$ has offset $\lambda q$ and span $\lambda$
for all $q\in\TbbbZ$.
\item
$(S|g_\lambda)^*(q)$ has offset
$\lambda\cdot\select_{g_\lambda(S)}(q)$
and span $\lambda$ for all $q\in\{1,\ldots,|g_\lambda(S)|\}$.
\end{itemize}

Using the rules to keep track of offsets and spans
enables us, at the bottom of a recursive
application of reductions,
to translate queries about sets correctly
to queries about their bit-vector representations
with the given offsets and spans.
So as not to clutter the description, this simple
translation will not be formulated explicitly.

\begin{figure}[!ht]
\begin{center}
\includegraphics{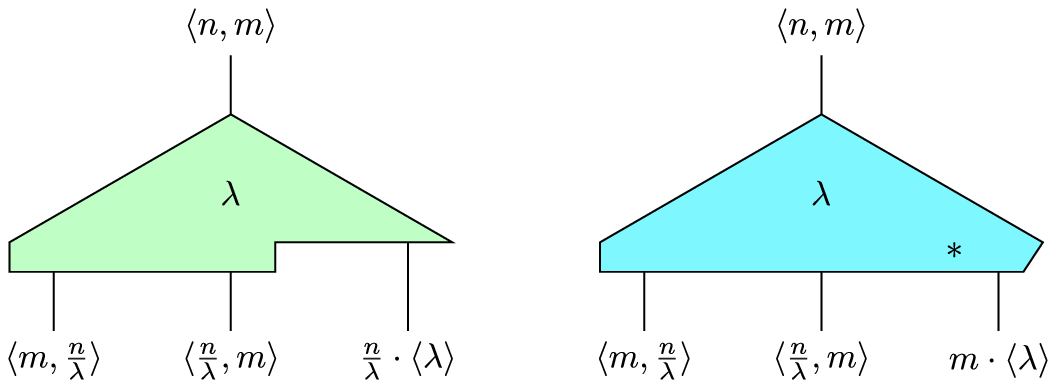}
\end{center}
\caption{The approximate effect of
the two combined reductions CR1 and
CR2 on $\Tangle{\mbox{span},\mbox{size}}$.}
\label{fig:effect}
\end{figure}

The effect of each type of combined reduction on
spans and sizes is depicted in Fig.~\ref{fig:effect}.
A pair of the form $\Tangle{n,m}$ indicates
that a set has span $n$ and size at most~$m$,
except that the rounding to integer values and
the occasional ${}+1$ were ignored.
When we refrain from bounding
the size of a set by anything better than its span,
$\Tangle{n}$ is used as an abbreviation
for $\Tangle{n,n}$.
If $S$ has offset $a$ and span $n$,
$(S|g_\lambda)(q)$ can be nonempty only if
$q$ belongs to the set
$\{g_\lambda(a),\ldots,g_\lambda(a+n-1)\}$
of size at most $\Tceil{\frac{n}{\lambda}}+1$,
which motivates the label
$\frac{n}{\lambda}\cdot\Tangle{\lambda}$
in the left subfigure.
In terms of the concrete data structure,
the expression $\frac{n}{\lambda}\cdot\Tangle{\lambda}$
should be thought of as indicating an array
of (approximately) $\frac{n}{\lambda}$
subordinate data structures,
each of which is for a set of span $\lambda$.
$(S|g)^*(q)$ is defined only for
$1\le q\le |g(S)|\le|S|$, which motivates the
label $m\cdot\Tangle{\lambda}$ in the right subfigure.

The new optimal rank-select index can be pieced together
with little effort from a constant number of instances
of the combined reductions
CR1 and CR2.
As a warm-up and to familiarize the reader with the
approach and the notation, we first develop a simpler
rank-select structure that occupies $\Theta(n)$ bits
for client sequences of $n$ bits.

\section{A Simplified $O(n)$-Bit Rank-Select Structure}
\label{sec:simplified}

\begin{figure}[!ht]
\begin{center}
\includegraphics{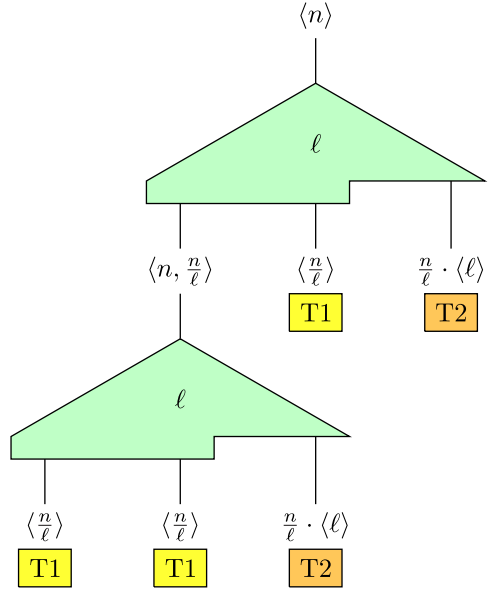}
\end{center}
\caption{A simplified rank-select structure that uses $O(n)$ bits.}
\label{fig:simplified}
\end{figure}

The simplified rank-select structure
is
best thought of
as the tree $T\Tsub S$
shown in Fig.~\ref{fig:simplified}
annotated with $\Tangle{\mbox{rank},\mbox{size}}$
pairs suitable for a client sequence of $n$ bits.
Each inner node in $T\Tsub S$ corresponds to
an instance of the composite reduction CR1
(for brevity, is a CR1-node) and is drawn with the
characteristic shape of that reduction.
Each leaf in $T\Tsub S$ corresponds to an
instance or an array of instances
of one of the variants of the
table-lookup method and is drawn as a rectangle labeled
with the name of the relevant variant.
During the execution of a query, each inner node in~$T\Tsub S$
applies its associated reduction
and each leaf
answers queries using its table-lookup variant.

The reductions in $T\Tsub S$ both use a parameter $\ell\in\TbbbN$.
Here and in the following, we choose $\ell=\Theta(\log n)$
such that the cost, in terms of time and space, of
the table needed by
table-lookup variant T2 with
$N=\ell$ is negligible.
In concrete terms, we take this to mean that
$2^\ell\ell\log\ell=O({n/\log n})$, which is certainly satisfied
if $\ell\le({1/2)}\log_2 n$.

The simplified rank-select structure is correct by construction
and trivially executes queries in constant time.
Let us analyze its storage requirements, for the time
being ignoring rounding issues and pretending that
the expressions for spans and sizes in Fig.~\ref{fig:effect}
are exact.
The first step is to verify that the
$\Tangle{\mbox{span},\mbox{size}}$ pairs in
Fig.~\ref{fig:simplified} have indeed be calculated
in accordance with Fig.~\ref{fig:effect}.
Each node in $T\Tsub S$ needs to store a constant number of
offsets, spans and other integers that allow it to access arrays
and bit sequences correctly.
Beyond this,
inner nodes
have no associated storage.
Each leaf that uses table-lookup variant T1
(is a T1-leaf, say)
stores a table of $\rank$ and $\select$
for a sequence of $\frac{n}{\ell}$ bits, which needs
$O(\frac{n}{\ell}\log n)=O(n)$ bits.
Similarly, each T2-leaf
stores an
array of $\frac{n}{\ell}$ sequences, each of
$\ell$ bits, again for a total of $O(n)$ bits.
Adding the $O({n/\log n})$ bits
occupied by a global table
for variant T2 and $O(\log n)$ bits for
a constant number of offsets, spans and other integers,
we arrive
at a grand total of $O(n)$ bits.
It is easy to see that the error incurred by the
approximation involved in
Fig.~\ref{fig:effect} amounts to less than a
constant factor
(this is because error terms bounded by constants
affect only quantities that are $\Omega(\log n)$),
so that the true number of bits
used by the simplified rank-select structure is
also $O(n)$.

\section{The New Optimal Rank-Select Index}

\begin{figure}[!ht]
\begin{center}
\includegraphics{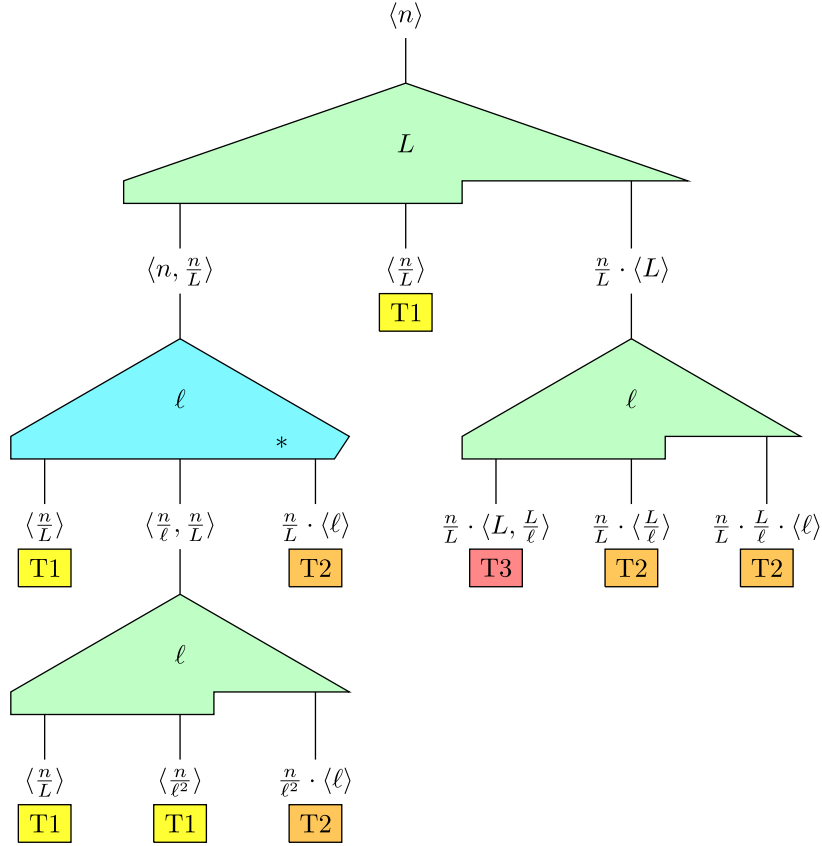}
\end{center}
\caption{The new optimal rank-select index.}
\label{fig:complete}
\end{figure}

The new optimal rank-select index has much in common with
the simplified rank-select structure of the previous section.
In order to achieve a better space bound, however,
the optimal index must comprise a few
additional reductions.
Its structure is given by the tree $T$
shown in Fig.~\ref{fig:complete}.
While most reductions use the parameter $\ell$
with $\ell=\Theta(\log n)$
introduced in Section~\ref{sec:simplified},
the reduction at the root of~$T$
employs a larger parameter $L\in\TbbbN$.
We choose $L=\Theta({{(\log n)^2}/{\log\log n}})$
as a multiple of $\ell$
such that the cost of
the table needed by
table-lookup variant T3
with $N=L+1$ and $M=1+{L/\ell}$ is negligible.
In concrete terms, we take this to mean that
$(2(L+1))^{2+{L/\ell}}\log L=O({n/\log n})$, which is ensured if
${L/\ell}\le({1/4}){{\log_2\!n}/{\log_2\!\log_2\! n}}$.

\subsection{Analysis of the Space Requirements}

As in the case of the simplified rank-select structure,
the optimal rank-select index is correct by construction and
trivially executes queries in constant time.
Because we now aim for a space bound of $o(n)$ bits,
we must pay special attention to the rightmost
leaf in the tree $T$ with its (approximately)
$\frac{n}{L}\cdot\frac{L}{\ell}=\frac{n}{\ell}$
sequences, each of (at most) $\ell$ bits.
As is not difficult to see, each of the relevant bit
sequences is a subsequence of the client sequence~$B$,
so that there is no need to store any data in
the rightmost leaf---it suffices to provide it with
constant-time access to arbitrary subsequences of
at most $\ell=\Theta(\log n)$ consecutive bits in~$B$,
which is precisely what the rank-select index is
allowed to rely on.

The remaining part of the analysis of the space
requirements
parallels
what was done in Section~\ref{sec:simplified}
for the simplified rank-select structure, and again
we can pretend that the expressions given in
Fig.~\ref{fig:effect} are exact---as a minor exception,
one should observe that
table-lookup variant~T3 is indeed
used with $N=L+1$ and $M=1+{L/\ell}$.
Again, the first step is to verify
that the $\Tangle{\mbox{span},\mbox{size}}$
pairs in Fig.~\ref{fig:complete} have been calculated
in accordance with Fig.~\ref{fig:effect}.

Each T1-leaf
holds tables of
$\rank$ and $\select$ for bit sequences of length
$\frac{n}{L}$ or $\frac{n}{\ell^2}$.
The number of bits needed for the tables is therefore
$O((\frac{n}{L}+\frac{n}{\ell^2})\log n)=O({n \log \log n/\log n})$.
The total length of the bit sequences stored in
T2-leaves
is $O((\frac{n}{L}+\frac{n}{\ell^2})(\ell+\frac{L}{\ell}))=
O(\frac{n}{L}\cdot\ell)=O({{n\log\log n}/{\log n}})$.
Finally, the single T3-leaf
needs space
for $\frac{n}{L}$ bit sequences, each of which represents
a tuple of $\frac{L}{\ell}$ integers of
$O(\log L)$
bits each, for a total of
$O(\frac{n}{L}\cdot\frac{L}{\ell}\cdot\log L)
 =O({{n\log\log n}/{\log n}})$ bits.
In summary, the number of bits occupied by the
rank-select index is
$O({{n\log\log n}/{\log n}})$.

\subsection{The Execution of Queries}
\label{subsec:queries}

In a concrete implementation of the rank-select index,
it is natural to represent each node in $T$ by
an instance of a suitable class that supports
operations $\rank$ and $\select$.
Assume that $U$ is such an instance that corresponds
to an inner node $u$ in $T$,
that $X$, $Y$ and $Z$
are the class instances that correspond to the
children of $u$, in the order from left to right,
and that $Z$ overloads the array-indexing syntax with an
operation that incorporates the implicit
offsets and spans as appropriate.
If $u$ is a CR1-node with parameter $\lambda$,
a combination of Equations~(\ref{eq:1})--(\ref{eq:4})
in Subsection~\ref{subsec:basic}
shows that $U$'s operations can be
realized according to the formulas
\[\begin{array}{r c l}
U.\rank(j) &=& \textbf{let }q=g_\lambda(j)\textbf{ in }
X.\select(Y.\rank(q-1)+1)+Z[q].\rank(j)\mbox{\quad and} \\[0.3em]
U.\select(k) &=& \textbf{let }q=Y.\select(X.\rank(k-1))\textbf{ in }
Z[q].\select(k-X.\select(Y.\rank(q-1)+1)).
\end{array}\]
The formulas
can clearly be expressed as
straight-line code.
If $u$ is a CR2-node,
the formulas become a little more complicated in
that, by Equation~(\ref{eq:5}),
$Z[q].\rank(j)$ must be replaced by
$Z[Y.\rank(q)].\rank(j)\cdot(Y.\rank(q)-Y.\rank(q-1))$
and $Z[q].\select(\ldots)$
by $Z[Y.rank(q)].\select(\ldots)$
(for $\select$, we always have
$Y.\rank(q)-Y.\rank(q-1)=1$).
The computation is still straight-line, but it
could be argued that it would be more natural
to replace the multiplication by a zero test
followed by a branch.
This is why claiming that our $\rank$ and $\select$
operations are straight-line involves a small
amount of ``cheating''.
Class instances that correspond to leaves in~$T$,
of course, realize their $\rank$ and $\select$
operations by a single access to a 1- or
2-dimensional array, handled in accordance with
the relevant offsets and spans.

We can improve
the implementation of $U.\select$
as follows:
If we compute $r=X.\rank(k-1)$ and then $q=Y.\select(r)$,
we can use the equality $Y.\rank(Y.\select(r)-1) = r-1$
to replace the subexpression $Y.\rank(q-1) + 1$ by~$r$,
thus eliminating one call of $Y.\rank$
by reusing~$r$.
If $u$ is a CR2-node,
since $Y.\rank(Y.\select(r)) = r$,
we can
avoid another call of $Y.\rank$
in the implementation of $U.\select$
by substituting
$Z[r].\select(\ldots)$ for
$Z[Y.\rank(q)].\select(\ldots)$.

\subsection{The Construction of the Index}
\label{subsec:construction}

In order to construct the optimal
rank-select index for a
client sequence $B$ of $n$ bits,
we interpret each node in the tree $T$
of Fig.~\ref{fig:complete} as a process
whose task is to communicate with adjacent
nodes and, in the case of leaves, to compute
and store a table or an array for use in
subsequent queries.
Each inner node in $T$ receives a stream of bits from
its parent
and sends streams of bits or integers
derived from its input stream to
its children.
An exception concerns the root of $T$, which receives
the client sequence $B$ in the stream that feeds the
overall construction and prepends a single \texttt{0}
to $B$ before processing $B$, which is now
considered to represent the same set as before,
but with offset~0 and span $n+1$.

In a preprocessing phase that proceeds top-down in
the tree and takes constant time, each node uses the
rules formulated in Section~\ref{subsec:combined} to
compute the spans and sizes of the (multi)sets that
it will handle and possibly other simple functions
of~$n$ that enable it to carry out the
relevant array accesses.
If the node is a leaf, it also acquires the space
needed to hold its table or array.

Consider an inner node $u$ in $T$ with
parameter $\lambda$ and assume that $u$'s input stream
is a bit-vector representation of a set $S$
whose offset $a$ is a multiple of~$\lambda$---by the
\texttt{0} prepended to the client sequence
as described above, this assumption, which we
call the \emph{alignment assumption}, is satisfied for
the root of~$T$.
If $u$ receives a sequence of bit-vector representations,
the assumption as well as the following
arguments should be applied
independently to each element in the sequence.
The task of $u$ is to send either
a stream of the elements of
$\rank_{g_\lambda\ms{S}}(\TbbbZ)$
(if $u$'s left child is a T3-leaf)
or a bit-vector representation of
this set with offset~0
(otherwise)
to its left child, a bit-vector representation of
$g_\lambda(S)$ with offset $g_\lambda(a)$
to its middle child, and bit-vector representations
of either $(S|g_\lambda)(q)$ with offset $\lambda q$
for $q=g_\lambda(a),g_\lambda(a)+1,\ldots$
or $(S|g_\lambda)^*(q)$ with offset
$\lambda\cdot\select_{g_\lambda(S)}(q)$
for $q=1,\ldots,|g(S)|$ to its right child.

The node $u$
processes its input stream in \emph{batches} of
$\lambda$ consecutive bits each, except that the
last batch may be smaller, in which case it
is filled up to size $\lambda$ with \texttt{0}s.
Note that by the alignment assumption,
a batch corresponds exactly
to $g_\lambda^{-1}(q)$ for some $q\in\TbbbZ$.
Before the processing of the first batch,
$u$ initializes a variable $s$ to~0
and sends the integer~0
(if $u$'s left child is a T3-leaf)
or a single \texttt{1} (otherwise)
to its left child.
The processing of a batch
begins by determining the number $k$ of
\texttt{1}s in the batch.
If $\lambda=\ell$, this is done in constant time
by lookup in a table whose construction time
($O(2^\ell)$)
and space requirements ($O(2^\ell\log\ell)$ bits)
are negligible.
If $\lambda=L$, $k$ is instead found by consulting
the table $L/\ell$ times and
summing the values found there.
If $k>0$, $u$ adds $k$ to $s$ and proceeds to send
the current value of $s$
(if $u$'s left child is a T3-leaf)
or $k-1$ \texttt{0}s
followed by a \texttt{1}
(otherwise) to its left child,
a \texttt{1} to its middle child and the
whole batch to its right child as
the next sequence element;
If $k=0$, $u$ instead sends nothing to its
left child, a \texttt{0} to its middle child and,
only if $u$ is a CR1-node, the whole batch
to its right child, again as
the next sequence element.

It is easy to see that the successive values of~$s$
are precisely the elements of
$\rank_{g_\lambda\ms{S}}(\TbbbZ)$,
so that what is sent by $u$
to its left child is either
a stream of the elements of
$\rank_{g_\lambda\ms{S}}(\TbbbZ)$ in sorted order
(if $u$'s left child is a T3-leaf)
or a bit-vector representation
with offset~0 of this set
(otherwise).
In particular, the alignment assumption is
satisfied at the left child $v$ of~$u$
unless $v$ is a leaf
(in which case no alignment is needed).
Similarly, the stream sent by $u$ to its middle
child is obviously the bit-vector representation
with offset $g_\lambda(a)$ of $g(S)$.
In the concrete tree of Fig.~\ref{fig:complete},
either $a=0$ (this is the case for all inner
nodes $u$ except the right child of the root)
or the middle child of $u$ is a leaf.
Thus the alignment assumption is satisfied at
the middle child $v$ of $u$ if $v$ is an inner node.
Finally, the fact that $u$ passes its input stream,
subdivided into batches, on to its right child
if $u$ is a CR1-node and does the same
but omitting the batches without
\texttt{1}s if it is a CR2-node
is easily seen to be in accordance with
the specification above.
If the right child $v$ of $u$ is not a leaf,
$u$ is the root of $T$, and the parameter $L$ of $u$
is a multiple of the parameter $\ell$ of $v$, so that the
alignment assumption is satisfied at~$v$.

Each T1-leaf in $T$
constructs and stores a table of $\rank$ and $\select$
for the bit stream that it receives, and
each T2-leaf receives a sequence of
bit streams and simply stores these in
successive cells of an array.
An exception concerns the rightmost leaf, which ignores
the bit stream that it receives and stores nothing.
Finally, the single T3-leaf, for each of the
(approximately) $\frac{n}{L}$ sets that it receives,
stores the concatenation of the
$\Tceil{\log_2(L+1)}$-bit binary representations
of the at most $1+L/\ell$ elements of the set
in the next cell of an array, precisely as
called for by table-lookup variant T3.

If we introduce a buffer of $L+\ell$ bits between each
pair of adjacent nodes in $T$ (except between the
single T3-leaf and its parent, where a buffer of
$1+{L/\ell}$
integers of $\Tceil{\log_2(L+1)}$ bits each
is suitable), we can repeatedly
execute a top-down sweep over $T$ in which the
root processes the next batch of $L$ bits of
the client sequence in
$O({L/\ell})$ time, thereby adding bits to its
outbuffers, and every other node in $T$ processes
as many integers or
complete batches of $\ell$ bits, each in constant
time, as available in its inbuffer, again adding
integers or bits to its outbuffers, if any.
It is easy to see that each sweep can be executed
in $O({L/\ell})$ time.
Then the whole process finishes in
$O({n/\ell})=O({n/{\log n}})$ time, and it
uses $O(L)$ bits in addition to what is needed to
store batches of $B$ and the finished data structure,
a total of
$O({{n\log\log n}/{\log n}})$ bits.
This concludes the proof of Theorem~\ref{thm:main}.

\subsection{Tuning the Tree of Reductions}

At the price of having to observe more closely what
happens in the individual nodes in the tree $T$
of reductions,
it is possible to modify $T$
in ways likely to
reduce the operation times and space
requirements of the data structure by
constant factors.

First, note that
table-lookup variant T1 can deal with a multiset
whose size and span are both bounded by
some $N\in\TbbbN$ as easily
as with a simple subset of $\{1,\ldots,N\}$.
If a set $S$ of integers is of span $n$ and size~$m$
and $\lambda\in\TbbbN$,
then $g_\lambda\ms{S}$ can be given span
$\Tceil{\frac{n}{\lambda}}+1$ and is of size~$m$.
As can be seen from Fig.~\ref{fig:effect},
this implies that every CR1-node
whose left and middle children are both T1-leaves
can be replaced by a BR1-node with a T1-leaf as
its left child and the former right child of
the CR1-node as its right child.
In other words, we can do away with
an instance of the reduction
BR2 and a T1-leaf.
This applies to the inner node of maximal depth
in Fig.~\ref{fig:complete} (and also in
Fig.~\ref{fig:simplified}).

Second, a similar modification can be carried out at the
right child~$u$ of the root in Fig.~\ref{fig:complete}.
Begin by dissolving the reduction CR1 into its
constituent parts, BR1 and BR2.
Then the BR2-node can be seen to be in charge of
answering queries about multisets whose span,
$1+\frac{L}{\ell}$,
is significantly smaller than their size, $L$.
For such multisets we introduce another
table-lookup variant:

\paragraph*{T4.}
If $S$ is a variable multiset of size at most $M$
consisting of integers
in $\{1,\ldots,N\}$ for some given $N\in\TbbbN$,
$S$ can be represented by the $N$-tuple
$(\rank_S(1),\ldots,\rank_S(N))$
of integers in $\{0,\ldots,M\}$.
Each of the $N$ integers can in turn be represented
in $\Tceil{\log_2 (M+1)}$ bits.
Since $2^{\Tceil{\log_2 (M+1)}}\le 2 M$,
we can answer $\rank$ and $\select$ queries about a multiset
stored in this way with a table of $O((2 M)^{N+1}\log M)$
bits that can be computed in $O((2 M)^{N+1})$ time.

\bigskip
We now replace the BR2-node and its children, a T3-leaf and a T2-leaf,
by a T4-leaf with the parameters $N=1+L/\ell$ and $M=L$.
For each of $O(\frac{n}{L})$ multisets,
the T4-leaf must store $O(\frac{L}{\ell})$ integers,
each of $O(\log L)$ bits, a total of
$O({{n\log\log n}/{\log n}})$ bits.
The size of the global table needed by the T4-leaf
and the time to compute it can be seen to be negligible.

Third, in some cases where a reduction is used to
answer $\rank$ and $\select$ queries,
it may be possible to
handle one of the two operations directly with
table lookup.
This speeds up the operation in question, but because
the relevant subtree can be tuned to deal only with
the other operation, additional benefits may accrue.
To apply this observation most effectively, we
also dissolve
the combined reduction CR1
at the root in Fig.~\ref{fig:complete} into
its constituent parts, BR1 and BR2.
Since the new BR2-node $u$
is in charge of queries about a multiset of large size ($n$)
but small span (approximately $\frac{n}{L}$),
it can answer $\rank$
queries with table-lookup variant T1---the
necessary table is of
$O(\frac{n}{L}\log n)=O({{n\log\log n}/{\log n}})$ bits.
As can be seen from Equation~(\ref{eq:4}), this means that $u$
will issue only $\rank$ queries to its left child~$v$
(the left child of the root in Fig.~\ref{fig:complete})
and only $\select$ queries to its right child
(the middle child of the root).
The discussion in Subsection~\ref{subsec:queries}
then implies that
$v$ will in turn issue only $\select$ queries to its left child
and $\rank$ queries to its middle and right children,
and this pattern repeats at $v$'s middle child.

We must still describe how to construct an
index modified as above.
Each change conceptually begins
by replacing a CR1-node
by a BR1-node and a BR2-node.
Consider a BR1-node $u$ with parameter $\lambda$
whose input stream is a bit-vector representation
of a set $S$ for which the alignment assumption holds.
As described for CR1-nodes in
Subsection~\ref{subsec:construction}, $u$ consumes its
input in batches of $\lambda$ consecutive bits,
counts the number $k$ of \texttt{1}s in each batch,
sums these counts in a variable $s$ that is
initialized to~$0$, and passes on every batch
to its right child.
Where the behavior of $u$ differs from that of
CR1-nodes
is in its communication with its left child.
Informally, the multiset $g_\lambda\ms{S}$ is represented
by the prefix sums of its multiplicities.
More precisely, $u$ transmits to its left child the value
of its variable $s$ both before the processing of the
first batch (when $s=0$) and after the processing of
each batch (not just after batches that contain
\texttt{1}s).

Consider next a BR2-node $u$.
By what was just described,
$u$ receives a nondecreasing sequence of integers
from its parent.
When receiving the first integer, which is 0,
it stores it in a variable $s$
and sends a \texttt{1} to its left child.
For each subsequent integer $s'$, $u$ does the
following:
If $s'>s$, $u$ computes $k=s'-s$,
sends $k-1$ \texttt{0}s
followed by a \texttt{1} to its left child,
sends a \texttt{1} to its right child,
and sets $s:=s'$.
If $s'=s$, $u$ just sends a \texttt{0} to its right child.

It can be seen that a BR1-node and a BR2-node,
combined as in Fig.~\ref{fig:combined1}, together
exhibit the behavior of a CR1-node
whose left child is not a T3-node.
On the other hand, if the left child of a BR1-node
is a T1-leaf or a T4-leaf instead of a BR2-node,
the leaf can easily construct its table, as
described earlier in this subsection,
from the stream that it receives.

\begin{figure}[!ht]
\begin{center}
\includegraphics{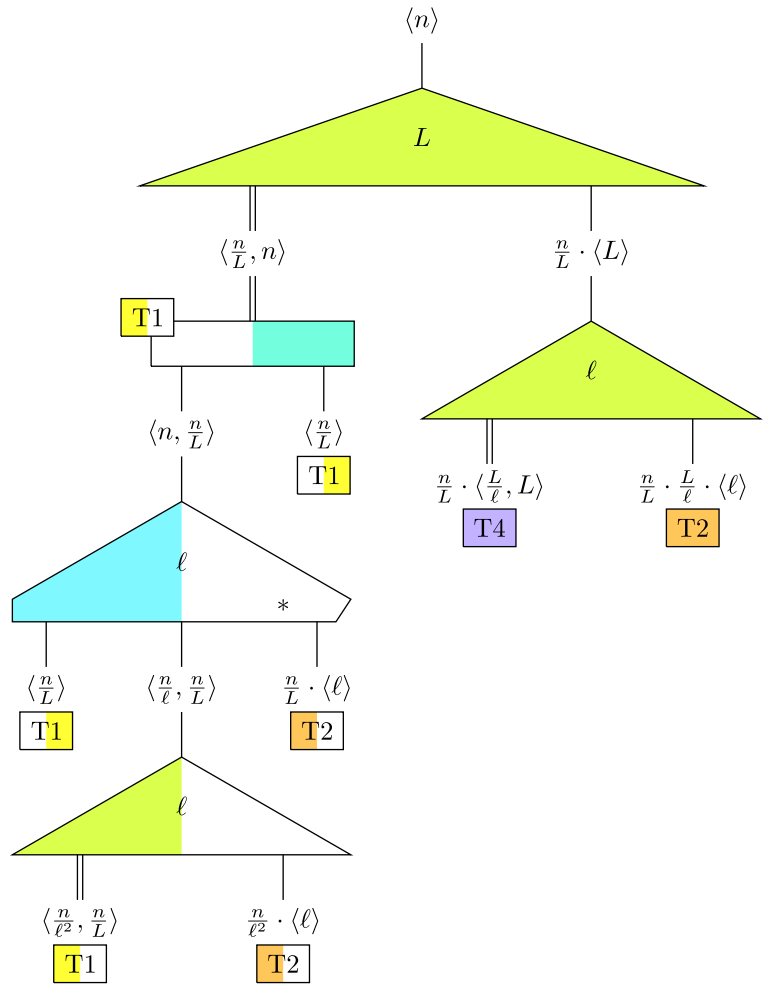}
\end{center}
\caption{A tuned rank-select index
with smaller tables and early completion of some queries.}
\label{fig:tuned}
\end{figure}

A tuned rank-select index that incorporates the
modifications discussed in this subsection is shown
in Fig.~\ref{fig:tuned}.
The notation $\Tangle{\mbox{span},\mbox{size}}$ is
now used also to characterize the salient properties
of multisets, but for a multiset the size may
exceed the span.
The operations $\rank$ and $\select$ were associated
arbitrarily with the left and right halves, respectively, of
(the graphical representation of)
each node, and if a node handles only queries of
one kind, only its corresponding half is shown colored
(by a rainbow color or gray).
The resulting savings, in terms of space, can be
estimated from a comparison of
Figs.\ \ref{fig:complete} and~\ref{fig:tuned}
(note here that relieving a T2-leaf of the burden of
answering either $\rank$ or $\select$ queries---but
not both---does not reduce the number of bits
that the leaf must store).

The evaluation of $\rank$ queries in the tuned index
follows a recipe that goes back to Jacobson~\cite{Jac88,Jac89}
and can hardly be improved upon:
The client sequence $B$ is partitioned into
\emph{superblocks}, each of these in turn is
partitioned into \emph{blocks}, and
to compute $\rank_B(j)$ we locate the superblock
and the simple block that contain the $j$th bit
(call these the \emph{target} superblock and block)
and add the number of \texttt{1}s before the
target superblock
(found in a \emph{first-level directory}),
the number of \texttt{1}s in the target superblock
but before the target block
(found in a \emph{second-level directory}),
and the number of \texttt{1}s in the target
block until and including the $j$th
bit of $B$, the latter quantity
being computed with table-lookup variant~T2.

\bibliographystyle{plain}
\bibliography{rank}

\end{document}